\documentclass[pdflatex,sn-mathphys-num,iicol]{sn-jnl}

\usepackage{graphicx}
\usepackage{multirow}
\usepackage{amsmath,amssymb,amsfonts}
\usepackage{amsthm}
\usepackage{mathrsfs}
\usepackage[title]{appendix}
\usepackage{xcolor}
\usepackage{textcomp}
\usepackage{manyfoot}
\usepackage{booktabs}
\usepackage{algorithm}
\usepackage{algorithmicx}
\usepackage{algpseudocode}
\usepackage{listings}
\usepackage{subcaption}
\usepackage{physics}
\usepackage{stmaryrd}
\usepackage{qcircuit}

\floatname{algorithm}{Benchmark Protocol}

\theoremstyle{thmstyleone}

\theoremstyle{thmstyletwo}

\theoremstyle{thmstylethree}

\begin{document}

\title[QWS]{Quantum WalkScore: Benchmarking Quantum Computers on the Graph Nodefinding Problem}

\author*[1]{\fnm{No\'e} \sur{Olivier}}\email{noe.olivier@thalesgroup.com}
\author[1]{\fnm{Michel} \sur{Nowak}}\email{michel.nowak@thalesgroup.com}

\affil*[1]{\orgdiv{CortAIx Labs}, \orgname{Thales Research and Technology}, \orgaddress{\city{Palaiseau}, \postcode{91120}, \country{France}}}

\abstract{
Recent advances in quantum computing hardware toward fault-tolerance have increased interest in evaluating near-term quantum platforms on application-relevant quantum algorithms.
In this work, we introduce Quantum WalkScore (QWS), a scalable application-oriented benchmark designed to assess the performance of NISQ and future fault-tolerant quantum computers in executing essential quantum routines --discrete-time quantum walks and quantum amplitude amplification-- to solve the graph nodefinding problem (marked-vertex search).
QWS quantifies performance by scoring the largest problem size for which a designated target node can be found with success probability above a defined threshold.
In addition to the complete protocol description, we provide example parameter-selection scenarios designed from noiseless simulation results and demonstrate QWS evaluation through experimental runs on multiple generations of IBM real quantum processors.}


\maketitle

\section{Introduction}\label{sec:introduction}

The numerous scientific and technological developments of the last decades have led to the emergence of quantum computers based on various promising technologies, architectures and computation paradigms.
While these systems are subject to continuous improvements on noise handling in order to eventually propose fault-tolerant quantum computation, the end-user community has shown a growing interest in the potential use of these computing capabilities and performance assessment.

As a consequence, numerous quantum computing benchmarking methods have been proposed to better grasp the level of maturity of these systems on multiple levels~\cite{wang2022sok}: physical, circuit, and application.
Hardware-level benchmarks such as gate set tomography~\cite{nielsen2021gate} help capture physical characteristics of hardware.
Circuit-level benchmarks include the Quantum Volume~\cite{quantumvolume} and randomized benchmarking approaches~\cite{proctor2017randomized,knill2008randomized}.
They aim to identify sources of error and evaluate the capability of quantum computers to successfully perform a series of operations on a group of qubits.
Application-oriented benchmarks focus on evaluating hardware performances in solving practical problems meaningful to the end-user.
With the latest developments of quantum devices leading to logical qubit operations demonstrations and the design of quantum algorithms for finance, optimization, physics simulation, and image analysis, among other application domains~\cite{dalzell2025quantum}, application-oriented benchmarks have emerged as necessary and important tools to help identify the practical benefits of each quantum computing platform.
In addition to application-centered individual benchmarks~\cite{martiel2021benchmarking, van2024extending,erbin2026many}, application-oriented benchmark suite initiatives combining fidelity estimation for meaningful algorithms~\cite{tomesh2022supermarq,donkers2022qpack}, end-user applications~\cite{barbaresco2025bacq,lubinski2023application}, and benchmarking frameworks~\cite{finvzgar2022quark, blume2020volumetric} have been designed.
However, the literature remains scarce in the use of promising quantum routines, especially discrete-time quantum walks, to benchmark quantum computers on graph application problems.

Quantum walks offer a promising method to finding solutions of optimization problems, providing a quadratic speedup in graph traversal problems on any graph~\cite{ambainis2020quadratic} compared to classical random walks.
In particular, multiple works have demonstrated its benefit in finding marked elements on one and two-dimensional graphs~\cite{ambainis2012search,giri2020lackadaisical,krovi2016quantum}, a task closely related to routing, supply-chain, or traveling salesman-like use-cases identified as natural applications for quantum walks.
The use of quantum walk-based optimization algorithms has also been investigated to address combinatorial problems in vehicle routing~\cite{bennett2021quantum} and portfolio optimization in finance~\cite{slate2021quantum,qu2024experimental} by leveraging continuous-time quantum walks as a generalization of QAOA's mixing operator.
Similarly, discrete-time quantum walks (DTQW) have been used as an algorithmic brick for the quantum Metropolis-Hastings algorithm and its application to the N-Queen problem~\cite{campos2023quantum} to show potential speedup for large-scale instances.
In addition, discrete-time quantum walks exhibit non-trivial behaviors that can be leveraged to simulate complex quantum systems, such as two-particle dynamics simulation~\cite{schreiber20122d}, and high energy physics through particle transport~\cite{olivier2026monte} or photon interaction cross sections~\cite{lee2025quantum}.

The coin-based DTQW algorithm acts on the system Hilbert space $\mathcal{H}=\mathcal{H}_p\otimes\mathcal{H}_c$, where $\mathcal{H}_p$ encodes the walker position and $\mathcal{H}_c$ encodes the internal degrees of freedom, i.e. the coin.
A single walk step is implemented by the unitary $U=S(I_p\otimes C)$.
The coin operator $C$ acts on a coin register to define and encode directions of propagation.
The shift operator $S$ acts on $\mathcal{H}$ and evolves the state of the position register according to the state of the coin qubits.
Practical implementations of the coin-based DTQW algorithm include the use of CNOT-based increment/decrement functions~\cite{douglas2009efficient,georgopoulos2021comparison,nandi2025robust} and QFT-based adder function~\cite{koch2022gate, razzoli2024efficient}.
Reported experiments are mostly conducted on noisy superconducting hardware for a single step on a two-dimensional lattice~\cite{acasiete2020implementation}, or multiple steps on the one-dimensional walk~\cite{wadhia2024cycle}.
The latter work hints to a possible benchmarking approach by evaluating quantum computers capability in simulating a $n$-step walk on a $n$-node cycle graph, and achieves a reasonably high fidelity up to $n=4$ on IBM's now deprecated 5-qubit QPU.

In the context of search applications, quantum walk algorithms can be associated to the quantum amplitude amplification (AA) routine to enhance the probability of finding marked elements in a graph~\cite{magniez2007search,sahu2024quantum}.
The quantum amplitude amplification algorithm proposed by Brassard~\cite{brassard2000quantum} has been applied to finance option pricing estimation~\cite{rebentrost2018quantum,kaneko2021quantum} and Monte Carlo simulations~\cite{montanaro2015quantum} as a subroutine within the quantum amplitude estimation algorithm of the same paper.
Similarly to quantum walks, although amplitude amplification is regarded as an essential routine to be incorporated in future practical quantum algorithms, related benchmarks in the literature mainly consist in estimating the fidelity of the quantum amplitude estimation algorithm implementation on hardware~\cite{lubinski2024quantum}.

\begin{figure}[h]
	\begin{subfigure}{.43\textwidth}
		\centering
		\includegraphics[width=0.99\linewidth]{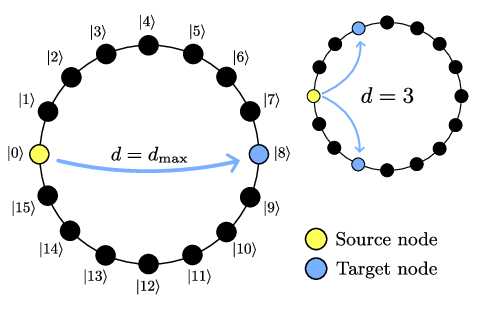}
		\caption{Cycle graph ($16$ nodes)}
	\end{subfigure}%
	\\
	\begin{subfigure}{.45\textwidth}
		\centering
		\includegraphics[width=0.99\linewidth]{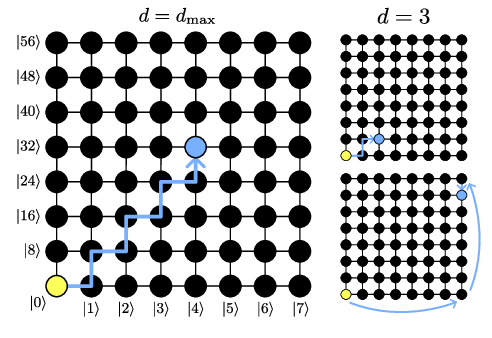}
		\caption{2D-torus graph ($64$ nodes)}
	\end{subfigure}
	\caption{\textsf{Representations of the graph nodefinding problem.} The cycle graphs (a) are generated for $n=4$, i.e. $4$ qubits, while the 2D-torus graphs (b) are generated for $n=3$, i.e. $3$ qubits per dimension. The target nodes (blue) are located at a distance $d=d_\text{max}$ (left) and $d<d_\text{max}$ (right) from the source node $\ket{0}$ (yellow).}
	\label{fig:graphs}
\end{figure}

In this work, we propose the design of Quantum WalkScore (QWS), a scalable application-oriented benchmark protocol based on solving the graph nodefinding problem using discrete-time quantum walks and amplitude amplification, two quantum routines especially promising for current and future fault-tolerant computations.
The graph nodefinding problem consists in finding a marked node in an undirected graph, given a specified initial node.
This problem serves as a fundamental building block across real-world applications such as telecommunication routing, database search, and complex network analysis where both DTQW and AA are expected to be applied.
For the purpose of benchmarking quantum computers of current and future times, the QWS protocol can be distinguished into two
different scores corresponding to two sets of graphs: cycle graphs and 2D-torus graphs illustrated in Fig~\ref{fig:graphs}.
One instance of a problem is defined by its size $(n,d)$ where $n$ is the number of qubits, and $d$ the distance between the initial and final nodes.
The benchmark therefore consists in iterative evaluations of the success probability of finding the target node after multiple steps of DTQW.
Incorporating the DTQW routine into the quantum amplitude amplification framework allows to increase the success probability via a reflection about the marked node, at the cost of a larger circuit depth.
The final score, QWS, corresponds to the largest problem size for which the success probability exceeds a given probability threshold $p^*(n,d)$ before a first failure.
Achieving a large score for this benchmark therefore requires a highly-performant QPU composed of high-fidelity qubits, operations, and long coherence times, along with carefully selected parameters: number of DTQW steps, number of AA iterations.

This paper is structured as follows.
In Section~\ref{sec:protocol}, we provide a complete description of the QWS protocol through the definition of the graph nodefinding problem, benchmark procedure, criteria for success, metrics and score evaluation. Additionally, we discuss the various choices made and experimental parameters recommendations.
In Section~\ref{sec:results}, we illustrate the QWS protocol through noiseless numerical simulations and discuss experimental results obtained on IBM quantum computers (Heron r2, Heron r3, and Nighthawk r1) by running the benchmark under two scenarios we have defined for parameters selection based on numerical results. 

\section{QWS protocol}\label{sec:protocol}

In this section we describe the proposed Quantum WalkScore (QWS) benchmark protocol and discuss the various choices made.
Similarly to other application-oriented benchmarks \cite{martiel2021benchmarking,erbin2026many}, the QWS is evaluated by  iteratively computing a given metric (i.e. probability of success) that is compared to a reference value.
The final score corresponds to the largest problem size for which the measured metric remains greater than the reference value.

The following paragraphs propose a detailed description of the studied problem, the structure of the benchmark, its parameters, and the score evaluation process.

\subsection{Description}\label{subsec:description}

\bmhead{Problem definition}

Let's define an undirected graph $G=(V,E)$ where $V=\{v_0, v_1, \ldots, v_m\}$ is the set of vertices (also called nodes) and $E$ the set of edges.
Two nodes $u, v \in V$ are connected if and only if $(u,v)\in E$.
Let's define a source node $v_0$ and a target node $v_t$.
\\

For the purpose of benchmarking quantum computers, we consider $G$ to be a cycle graph or a 2D-torus graph (Fig~\ref{fig:graphs}). These graphs are easily implemented and scalable, making them relevant graphs for benchmarking current and future hardware.

The graph nodefinding problem consists in finding the target node $v_t$ from the source node $v_0$ by moving along edges of the graph.

We define the distance between two nodes of a graph $G$ as the number of edges in the shortest path between them.
Let $d\in\mathbb{N}^*$ be the distance between $v_0$ and $v_t$, and let $n\in \mathbb{N}^*$ be a positive integer.
Let $d_\text{max}$ be the maximum distance between two nodes of a given graph.
We define the criteria for success, corresponding to the cycle graphs and 2D-torus graphs respectively, according to a threshold probability:
\\

$T^{(n,d)}_\text{cycle}$:
\textit{Find the target node, located at a distance $d$ from $v_0$ in the cycle graph of $2^n$ nodes, with a probability} $p\geq p^*_\text{cycle}(n,d)$. \\

$T^{(n,d)}_\text{torus}$:
\textit{Find the target node, located at a distance $d$ from $v_0$ in the 2D-torus graph of $2^n\times 2^n$ nodes, with a probability} $p\geq p^*_\text{torus}(n,d)$.

\begin{figure}[h]
	\centering
	\[
	\Qcircuit @C=0.4em @R=0.4em {
		& & & & & \ustick{\times \textsf{ num-aa} \text{ steps}} & & & & & \\
		\lstick{\ket{x_1}} & \qw & \multigate{3}{U(w)} & \qw & \multigate{1}{S_\chi} & \multigate{3}{U(w)^{-1}} & \multigate{1}{S_0} & \multigate{3}{U(w)} & \qw & \qw \\
		\lstick{\ket{x_0}} & \qw & \ghost{U(w)} & \qw & \ghost{S_\chi} & \ghost{U(w)^{-1}} & \ghost{S_0} & \ghost{U(w)} & \qw & \qw \\
		\lstick{\ket{c_1}} & \qw & \ghost{U(w)} & \qw & \qw & \ghost{U(w)^{-1}} & \qw & \ghost{U(w)} & \qw & \qw \\
		\lstick{\ket{c_0}} & \qw & \ghost{U(w)} & \qw & \qw & \ghost{U(w)^{-1}} & \qw &\ghost{U(w)} & \qw & \qw
		\gategroup{2}{5}{5}{8}{.8em}{--}
	}
	\]
	\caption{Quantum circuit implementation of the $w$-step discrete-time quantum walk $U(w)$ integrated within the quantum amplitude amplification framework from \cite{brassard2000quantum}. The operators $S_\chi$ and $S_0$ change the sign of the amplitude if and only if the states are $\ket{v_t}_p$ and $\ket{0}_p$ respectively. The position state is given by $\ket{x_1x_0}_p$ and the coin state by $\ket{c_1c_0}_c$. This circuit constructs $\ket{\psi_{w,aa}}_p$ when applied to $\ket{v_0}_p\ket{0}_c$.}
	\label{fig:circuitDTQWAA}
\end{figure}
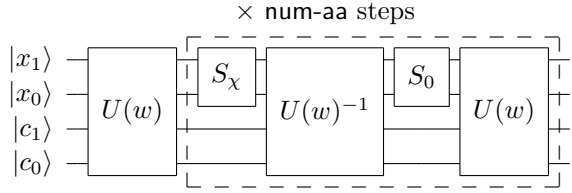

\bmhead{General structure of the benchmark}
Here, the QWS protocol is defined to be applied to gate-based quantum computers.
Its general structure can be decomposed as follows, starting from the minimum problem size $(n,d)=(2,1)$:
\begin{enumerate}
	\item Given $n,d \in \mathbb{N}^*$, construct the graph $G$ of $2^n$ nodes (for cycle graphs), resp. $2^n\times 2^n$ nodes (for 2D-torus graphs), and select the target node from the set of nodes located at a distance $d$ from the source node $v_0$. \\
	
	\item Initialize the $n$-qubit system (resp. $2n$-qubit system), in the state $\ket{v_0}_p$ and evolve it according to the \textsf{num-w}-step discrete-time quantum walk algorithm to construct $\ket{\psi_w}_p$. \\
	The value of \textsf{num-w} must be such that $\textsf{num-w}\geq d$ and $\textsf{num-w}\equiv d \pmod 2$ to ensure that the walker can reach the target node. \\
	
	\item Amplify the amplitude of the state corresponding to the target node via \textsf{num-aa} iterations of the quantum amplitude amplification routine to construct $\ket{\psi_{w,aa}}_p$. The corresponding quantum circuit is represented in Fig~\ref{fig:circuitDTQWAA}. \\
	
	\item Measure the success probability $p_\text{exp}(n,d) = |\bra{\psi_{w,aa}}\ket{v_t}|^2$ and compare the resulting metric to the threshold probability $p^*(n,d)$ according to criterion $T^{(n,d)}_G$. If the test is successful, consider the next problem size.\\
	
	\item The Quantum WalkScore is calculated as the maximum problem size for which the criterion remains successful before first failure. \\
\end{enumerate}

The metric of interest for a single experiment is the mean probability of success $p(n,d)$, i.e. the mean probability of measuring the target node's state $\ket{v_t}$. The calculation of this metric is detailed in a following paragraph.
The QWS is calculated from a series of such experiments, starting from the minimum problem size $(2,1)$ (for both graph families).
Let $\widetilde{n}, \widetilde{d}\in\mathbb{N}^*$ be defined such that
\begin{equation*}
\begin{cases}
\forall n\leq\widetilde{n}, \ \forall d\leq\widetilde{d}, \ T^{(n,d)}_G \text{ successful} \\
T^{(\widetilde{n},\widetilde{d})+1}_G \text{ unsuccessful}
\end{cases}
\end{equation*}
where the notation $(n,d)+1$ corresponds to the next possible pair of integers such that $d\leq d_\text{max}$.
In practice, the next problem size corresponds to increasing the distance between $v_0$ and $v_t$ by $1$ while keeping $n$ constant until we reach $(n, d_\text{max})$, in which case the next problem size becomes $(n+1, 1)$.
Note that $d_\text{max} = 2^{n-1}$ for the cycle graph and $d_\text{max} = 2^n$ for the 2D-torus graph.
The Quantum WalkScore is given by the following formula:
\begin{align}
\text{QWS}_\text{cycle} &= 2^{\widetilde{n}-1} + \widetilde{d} \label{eq:qwscycle}\\
\text{QWS}_\text{torus} &= 2^{\widetilde{n}} + \widetilde{d} \label{eq:qwstorus}
\end{align}

The goal of the proposed benchmark is twofold.
On the one hand, it must evaluate the capabilities of a given quantum computer to solve the graph nodefinding problem.
Hence, the scores in Eq.~(\ref{eq:qwscycle}) and Eq.~(\ref{eq:qwstorus}) are defined such that their value increases continuously with the size of the problem. This also allows for any end-user to easily identify the largest problem size solved successfully before a first test failure.

On the other hand, the benchmark must ensure that the criterion is considered successful if, and only if, the solution is significantly better than a random guess.
This aspect is taken into account with the definition of the threshold probability $p^*(n,d)$.

\bmhead{Threshold probability $p^*(n,d)$}
The threshold success probability $p^*(n,d)$ is designed to ensure that any measured success probability exceeding this value cannot be attributed to random guessing. For both graph families, it is defined as
\begin{equation}
	p^*(n,d) =
	\begin{aligned}
	\begin{cases}
	1\slash 3 & \forall d\leq2, \\
	1\slash 5 + 1\slash 2^d & \forall d>2
	\end{cases}
	\end{aligned}
	\label{eq:thresh}
\end{equation}

The value of $p^*(n,d)$ is used as the minimal success probability required to declare the test $T^{(n,d)}$ successful. \\
\\
\textbf{Metric and score evaluations. }
The threshold success probability is compared to a success probability metric to validate the test $T^{(n,d)}$.

We call problem instance a problem specified by its size $(n,d)$ and a selected target node $t$.
For each problem size, the success probability metric is estimated from experimental results obtained by running $N_\text{batch}=4$ batches of $N_\text{shots}=1000$ shots for each problem instance.
The stabilized metric $p(n,d)$ for the problem size $(n,d)$ is then computed by averaging the measured success probabilities across all instances and batches as:
\begin{equation}
	\label{eq:metric}
	p(n,d) = \frac{1}{N_\text{batch}|\mathcal{T}|}\sum_{t\in\mathcal{T}} \sum_{k=1}^{N_\text{batch}} p_\text{exp,k}(n,d,\text{target=t})
\end{equation}
where $|\mathcal{T}|$ is the number of selected target nodes (i.e. number of problem instances).
In particular for cycle graphs, all possible targets are selected, i.e. $|\mathcal{T}| = 2$, while for 2D-torus graphs, we select four different target nodes, i.e. $|\mathcal{T}| = 4$.
This number is chosen as a trade-off between representativeness of possible target nodes and computational cost of the benchmark evaluation.
Note that these values are true when $d<d_\text{max}$, and $|\mathcal{T}| = 1$ when $d=d_\text{max}$.

The score QWS is then computed according to Eqs.~(\ref{eq:qwscycle}) and (\ref{eq:qwstorus}), where $(\widetilde{n},\widetilde{d})$ correponds to the last problem size for which all tests succeed before the first failure.

\bmhead{Protocol parameters specifications}

In order to ensure the validity of experimental results, let's define the scope of parameters choice for the end-user.
\begin{itemize}
	\item \textit{Problem sizes } The minimum problem size is $(n,d)=(2,1)$ and the maximum problem size is at the hand of the end-user. \\
	
	\item \textit{Source node } Since both cycle and 2D-torus graphs are symmetric and have periodic boundary conditions, any node can be selected as the source node. By default: node $0$. \\
	
	\item \textit{Target nodes/Instances selection }	For cycle graphs: find the two nodes located at a distance $d$ from the source node. 
	For 2D-torus graphs: select at least four nodes located at a distance $d$ from the source node. When $d=d_\text{max}$, the unique possible target node is considered.\\
	
	\item \textit{Number of shots } It is required to have a total of $N_\text{shots}^\text{tot}=4000$ shots which can be dispatched into multiple batches of no less than $1000$ shots. We recommend to dispatch runs into $N_\text{batch}=4$ batches of $N_\text{shots}=1000$ shots per problem instance. \\
	
	\item \textit{Number of circuits } For cycle graphs we consider 2 instances (target nodes), thus a recommended number of $8$ circuit runs per problem size.
	For 2D-torus graphs we consider 4 instances, thus a recommended number of $16$ circuit runs per problem size. \\
	
	\item \textit{Threshold probability } The threshold probability is defined by Eq.~(\ref{eq:thresh}) and is compared to the metric experimental value defined by Eq.~(\ref{eq:metric}). \\	

	\item \textit{Discrete-time quantum walk } It is required to apply $\textsf{num-w}\geq d$ steps of the DTQW routine, based on a Hadamard coin and symmetric coin initial state: $SH\ket{0}_c$ (cycle graph), $SH\otimes I \ket{00}_c$ (2D-torus graph). The exact value of $\textsf{num-w}$ is left to be fine-tuned by the end-user.\\
	
	\item \textit{Amplitude amplification } One can apply $\textsf{num-aa}\geq 0$ iterations of the amplitude amplification routine. The value of $\textsf{num-aa}$ is left to be fine-tuned by the end-user. \\
	
	\item \textit{Error mitigation/correction } Error mitigation techniques or error correction codes can be used by the end-user to maximize the QWS value as long as they are mentioned in the publication of results.
\end{itemize}

\bmhead{Reporting results}\label{sec:reporting}

When publishing experimental results for the QWS protocol, it is recommended to provide comprehensive information on selected experimental parameters for reproducibility.
In particular, reports should include target nodes selection process, algorithm parameters selection process, circuits execution parameters (number of shots, number of batches), and error mitigation methods if any.

\subsection{Discussion}\label{subsec:discussion}

Let's discuss the various choices made in the protocol proposal.

\bmhead{Benchmark relevance}

The goal of the proposed QWS protocol is to evaluate the capability of quantum hardware to solve a graph nodefinding problem (related to various optimization problems on graphs) by using relevant and promising quantum routines.

Let's review how this proposal accounts for important aspects a relevant benchmark must answer~\cite{proctor2025benchmarking}.

\begin{enumerate}
	\item \textbf{Representativeness } Solving the graph nodefinding problem demonstrates the ability to efficiently search within structured and complex environments.
	This problem serves as a fundamental building block across diverse application domains in real-world complex network analysis, such as logistics (e.g. supply chains), telecommunications routing and navigation, infrastructure network threat identification, and database search~\cite{childs2004spatial}.
	
	In addition, the QWS protocol emphasizes DTQW and AA~\cite{brassard2000quantum}, two promising quantum routines that have demonstrated quadratic speedups over classical counterparts in terms of query or time complexity~\cite{ambainis2020quadratic,grover1996fast}. These algorithmic routines are widely regarded as essential building blocks for the development of practical quantum algorithms in a range of problem domains~\cite{montanaro2015quantum}. \\
	
	\item \textbf{Scalability } The scalability of the benchmark is controlled by tuning the size of graph $n$ and the distance parameter $d$.
	The selected graph families (i.e. cycle and 2D-torus graphs) have highly symmetric and periodic topologies, facilitating both algorithmic implementation (by enabling uniform circuit patterns) and mapping onto quantum hardware architectures with limited qubit connectivity.
	Note that this benchmark can in principle be adapted to more complex graphs that would better represent practical applications.
	Starting from small instances, such as those of the $(2,1)$ graph, the benchmark ensures a gradual increase in problem complexity that enables incremental evaluation of NISQ devices as hardware capabilities improve.
	
	Although the number of required qubits scales linearly with the size of the problem, implementing quantum algorithms based on DTQW and AA remains resource-intensive. In particular, circuit depth is strongly influenced by the choice of AA iterations \textsf{num-aa} and DTQW steps \textsf{num-w}.
	The total number of operations scales polynomially with the size of the problem (as shown in Fig~\ref{fig:depthsuff}), which presents challenges for current hardware due to error accumulation and decoherence.
	Hence, this benchmark yields a scalable and quantitative measure that can differentiate hardware performance across NISQ and, ultimately, FTQC devices and supports meaningful comparative assessments as quantum computers mature. \\
	
	\item \textbf{Equity } The specifications of this benchmark, including the choice of graph architectures and algorithm design, are devised to be applicable, in principle, to any universal quantum computer and all paradigms supporting gate-based algorithm implementations. Experimental parameters ranges are standardized in order to ensure statistically consistent results and fair comparison of hardware performance.

	Given that this benchmark is application-oriented, we intentionally leave the optimization of both $\textsf{num-aa}$ and $\textsf{num-w}$ parameters, as well as decisions regarding the use of error mitigation and correction techniques, to the end-user.
	This flexibility satisfies the diversity of application requirements and hardware capabilities. \\
	
	\item \textbf{Reproducibility } The score is entirely determined by two sets of parameters: the number of walk steps and AA iterations for each problem size, in addition to potential error mitigation methods.
	Adopting the recommended reporting practice, which requires that all experimental parameters values and mitigation techniques be explicitly documented, also ensures reproducibility of the results. \\
	
	\item \textbf{Transparency } The QWS is an informative performance indicator that enables direct and fair comparison of hardware performances for the proposed problem. It is explicitly defined as the largest problem size successfully solved before failure, a meaningful score that is readily interpretable even by non-expert end-users. \\
	
	\item \textbf{Verifiability } The resulting score can be easily verified by comparing experimental data to the threshold probability of success $p^*(n,d)$.
\end{enumerate}

\bmhead{Choice of graph families}

For this benchmark, we have selected two families of graphs: cycle graphs and 2D-torus graphs, chosen to balance simplicity, scalability, and hardware applicability.

Cycle graphs serve as a one-dimensional baseline problem, allowing straightforward evaluation of the metric on a simple topology.
Each node of the graph is connected to exactly two neighbors, translating into a qubit connectivity of two (ideally, which is expected to be supported by all gate-based quantum hardware).
This ensures that the benchmark can be successfully applied across existing NISQ devices and FTQC architectures.

In contrast, 2D-torus graphs represent a higher-dimensional problem, more relevant for estimating quantum device capabilities on more complex problems and quantum algorithms that better reflect the complexity of real-world applications.
Here, each node is connected to four neighbors due to periodic boundary conditions, which we employ to facilitate the implementation of quantum algorithms for metric evaluation.
While the ideal qubit connectivity of four is more demanding, it is not required to encode the problem at the cost of additional compilation overhead. This remains a practical target for near-future devices such as IBM's Nighthawk r1 device.

Additionally, the DTQW algorithm, which forms the basis for this benchmark, has been extensively studied on both cycle and torus graphs, providing well-documented theoretical and empirical bases.

Looking ahead, considering more complex graph structures involving higher dimensions and variable node degrees could generalize this benchmark to better capture constraints present in practical problems.

\bmhead{Choice of quantum routines}

In this protocol, we evaluate the capability of quantum computers to implement two key quantum routines: the discrete-time quantum walk (DTQW) and quantum amplitude amplification (AA).

The motivation for combining these algorithms is twofold.
First, DTQW provides an efficient algorithm for traversing a graph, offering a quadratic speedup in the propagation of the walker over a classical random walk~\cite{ambainis2020quadratic}.
Meanwhile, the AA routine provides a quadratic speedup in the number of queries compared to classical Monte Carlo sampling~\cite{montanaro2015quantum}.
Together, these algorithms form core components of various quantum algorithms and enable meeting the performance criterion of the successful test $T^{(n,d)}_G$.

Second, the DTQW algorithm has been extensively studied and experimentally implemented on quantum computers~\cite{douglas2009efficient,georgopoulos2021comparison,nandi2025robust,koch2022gate,razzoli2024efficient,acasiete2020implementation,wadhia2024cycle,magniez2007search}.
Importantly, when combined with AA, the DTQW circuit is repeated $O(\textsf{num-aa})$ times before measurement, and typically results in a circuit depth that scales as $O(nd)$ for our purpose.

In practice, the number of quantum walk steps $\textsf{num-w}$ must be greater than the distance $d$ to reach the target node.
The choice for a Hadamard coin and symmetric coin initial state is done to consider an unbiased walk across the graph.
This allows for an easy validation of the quantum walk implementation of a given hardware due to the theoretically symmetric distribution of probabilities across directions of each graph.
The number of AA steps $\textsf{num-aa}$ is to be chosen by the end-user.
Since increasing $\textsf{num-aa}$ significantly affects the circuit depth, one may prefer to optimize these two hyperparameters to limit qubit operations, subject to error propagation.

Ultimately, a successful execution depends not only on the choice of algorithmic parameters but also on hardware characteristics, including qubit and gate fidelities, coherence times, qubits connectivity, operation speed, and state preparation and measurement (SPAM) errors.
This interplay of algorithmic and hardware challenges makes the combination of DTQW and AA a suitable basis for our performance benchmark for quantum devices.

\bmhead{Choice of threshold probability $p^*(n,d)$}

The threshold probability defined in Eq.~(\ref{eq:thresh}) ensures that passing the test corresponds to succesfully solving the graph nodefinding problem rather than succeeding by random chance.
On one hand, the offset term $1\slash 5$ ensures that the threshold consistently exceeds the success probability achievable by random guessing with a significant margin (i.e. $p^*(n,d)\geq 150\% p_\text{rdm}(n,d)$) as illustrated in Fig.~\ref{fig:perfkingston}.
On the other hand, this value provides a constant performance objective, independent from the problem size, that guarantees the end-user a reasonable algorithmic performance for the stated problem.
Consequently, exceeding the threshold reliably indicates genuine algorithmic success.
This lower bound on threshold probability also ensures that the number of samples required to statistically validate the test remains bounded and independent of $n$, as discussed below, contributing to the benchmark's scalability.

Furthermore, by depending solely on the distance parameter $d$, rather than the graph size $n$, the threshold reflects the primary factor influencing problem complexity.
The exponential decay term $1\slash 2^d$ relaxes the success criterion as $d$ increases, homogenizing difficulty across problem scales.

Finally, the choice of an upper bound $p^*(n,d)=1\slash 3$ for instances with $d\leq 2$ allows for small problem instances to be solved without amplitude amplification when real hardware have sufficiently low error rates.
In contrast, achieving experimental success probabilities exceeding $p^*(n,d)$ on larger-scale problems typically necessitates employing quantum amplitude amplification.

\bmhead{Theoretical guarantees}

The existence of algorithmic parameters \textsf{num-w} and \textsf{num-aa} that allows theoretical success of the benchmark's criteria for any problem size is a result of the AA routine~\cite{brassard2000quantum}.

The DTQW algorithm, assuming $\textsf{num-w}\geq d$ and $\textsf{num-w}\equiv d \pmod 2$, povides an initial probability $a$ of measuring the target state $\ket{v_t}_p$ such that $a>0$. The Theorem 2 proposed by Brassard et al.~\cite{brassard2000quantum} guarantees the existence of an integer $m$ such that, by computing $\textsf{num-aa}=m$ iterations of the AA routine as proposed in Fig~\ref{fig:circuitDTQWAA} and measuring the system, the outcome is a success with probability at least $\max(1-a,a)$.
In particular, this value is given by $m=\lfloor\pi\slash 4\theta_a\rfloor$ where $\sin[2](\theta_a)=a$.

Since $p^*(n,d)\leq 1\slash 3 < \max(1-a,a)$, this demonstrates achievable theoretical success for all problem sizes on both graph families.

\bmhead{Experimental parameters}

In this paragraph, we address the quantitative choices of experimental parameters, with a focus on $N_\text{shots}$ and $N_\text{batch}$.

To estimate the success probability for a given problem instance with sufficient statistical confidence relative to $p^*(n,d)$, the number of shots must be large enough to achieve an acceptable margin of error.
Since the threshold probability $p^*(n,d)$ has a fixed lower bound independent of the problem size, the required number of shots need not increase with the problem size to maintain statistical rigor.
The choice of $N_\text{shots}^\text{tot} = N_\text{shots}\times N_\text{batch}$ reflects a trade-off between the desired confidence level in determining whether the metric exceeds the threshold and the practical availability of quantum hardware resources.
Increasing shots reduces statistical uncertainty, improving the precision and reliability of the estimated success probability, but at the cost of longer experimental runtime, higher resource consumption, and higher financial costs.

Specifically, the value of $N_\text{shots}^\text{tot}$ is chosen to estimate the success probability for a given problem instance (i.e. fixed graph size and target node) with a margin of error $\varepsilon = 1\%$ at a one-tailed $90\%$ confidence level, which are considered acceptable for the benchmark's needs.

We use the sample size formula for a binomial proportion
\begin{equation*}
	N_\text{shots}^\text{tot} = \frac{z_{0.10}^2 p(1-p)}{\varepsilon^2}
\end{equation*}
where $z_{0.10}\approx 1.28$ is the critical value for the one-sided test, and $p$ is the success probability.
Substituting values, considering $p=0.5$ in the worst case, gives $N_\text{shots}^\text{tot}\approx 4000$.
This corresponds approximately to a three-standard-deviation ($3\sigma$) margin of error of $2.37\%$.
This level of precision is sufficient to confidently distinguish successful tests from failures for practical benchmarking purposes.
Increasing the confidence level to 95\%, or tightening the margin of error to $\varepsilon=0.5\%$, would make the benchmark even more rigorous (i.e. reduced variance in the metric estimate) at the cost of significant computational cost (e.g. $2$ to $4$ times the total number of shots $N_\text{shots}^\text{tot}$), which may be prohibitive depending on the QPU service provider. \\

\begin{figure}[H]
	\centering
	\includegraphics[width=0.46\textwidth]{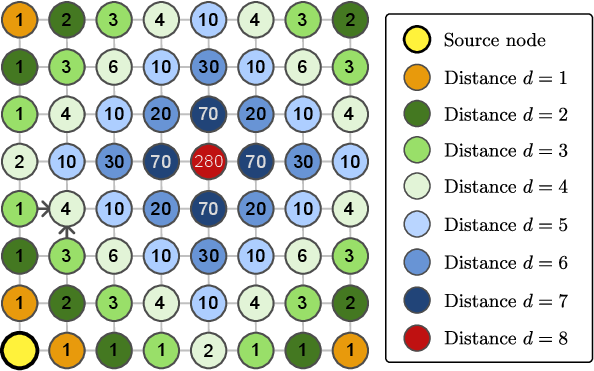}
	\caption{Representation of the number of possible paths to each target node for a $d$-step random walk on a $64$-node 2D-torus graph.}
	\label{fig:torusPaths}
\end{figure}

For the purpose of the QWS benchmark, we recommend distributing these $4000$ shots into $N_\text{batch}=4$ batches of $N_\text{shots}=1000$ shots for a given problem instance (fixed graph size $n$, distance $d$, and target node).
This batching improves hardware reliability in the results (e.g allowing better scheduling and intermediate hardware recalibration), improves noise detection ability (through multiple independent estimates), and ensures metric stability.
For cycle graphs, all two possible target nodes are considered, while for 2D-torus graphs, a fixed subset of four target nodes is selected.
This results in evaluating $8$ circuits per problem size for cycle graphs, and $16$ circuits for 2D-torus graphs, from which the metric $p(n,d)$ is computed (see Eq.~(\ref{eq:metric})).
Limiting the number of target nodes balances manageable computational cost with sufficient diversity of evaluated instances to improve robustness of the metric and fairly assess algorithm and hardware performance.
Although heuristic selection of target nodes that are potentially favored (see Fig~\ref{fig:torusPaths}) by the quantum walk algorithm might be considered for 2D-torus graphs, the necessary use of amplitude amplification to achieve high success probabilities limits any advantage from such optimization in reducing algorithmic resource requirements as shown in Figs~\ref{fig:torusperf} and~\ref{fig:torusresources}.

\section{QWS results on IBM quantum computers} \label{sec:results}

In this section, we discuss a first implementation of the QWS benchmark protocol on multiple IBM real quantum hardware.
In particular, we introduce some example scenarios where algorithm parameters \textsf{num-w} and \textsf{num-aa} are tuned based on simulated algorithmic performances on an ideal quantum computer.
Then, we evaluate and discuss the performance of real IBM quantum computers for these scenarios in order to provide a first estimate of their Quantum WalkScore.

\subsection{Experimental hardware specifications}

The QWS protocol is implemented using qiskit 2.3.1~\cite{javadi2024quantum}, qiskit-aer 0.17.2 for noiseless simulations, and QiskitRuntimeService from qiskit-ibm-runtime 0.46.1 to access IBM's quantum computers and emulators.

We choose to demonstrate our benchmark protocol on three different QPU detailed in Table~\ref{tab:qpu}, such that we evaluate three generations of IBM's hardware: Heron r2, Heron r3, and Nighthawk r1.
The reported specifications correspond to those of the day of experiments.

\begin{table*}[h]
	\centering
	\caption{Specifications of IBM hardware evaluated on the QWS benchmark.}\label{tab:qpu}%
	\begin{tabular}[width=\textwidth]{@{}llccccc@{}}
		\toprule
		Hardware & Generation  & T1 ($\mu s$) & T2 ($\mu s$) & 2Q-error & Readout error & CLOPS \\
		\midrule
		ibm\_kingston	& Heron r2		& 176	& 111	& $2.25\times 10^{-3}$	& $9.89\times 10^{-3}$	& 340K  \\
		ibm\_marrakesh	& Heron r2		& 175	& 77	& $2.73\times 10^{-3}$	& $12.2\times 10^{-3}$	& 300K  \\
		ibm\_pittsburgh	& Heron r3		& 257	& 277	& $1.63\times 10^{-3}$	& $4.88\times 10^{-3}$	& 330K  \\
		ibm\_miami		& Nighthawk r1	& 351	& 231	& $2.78\times 10^{-3}$	& $2.14\times 10^{-3}$	& 24K   \\
		\botrule
	\end{tabular}
\end{table*}

\subsection{Methods}

\bmhead{Scenarios and algorithm parameters}

We propose two scenarios for the design of the algorithm parameters $(\textsf{num-w}, \textsf{num-aa})$ and performance evaluation:
\begin{itemize}
	\item Scenario \textit{suff-perf}: $(\textsf{num-w}_\text{suff}, \textsf{num-aa}_\text{suff})$ 
	\item Scenario \textit{best-perf}: $(\textsf{num-w}_\text{best}, \textsf{num-aa}_\text{best})$. 
\end{itemize}

Let $\mathbb{N}_W\subset \mathbb{N}$ (resp. $\mathbb{N}_{AA}\subset\mathbb{N}$) be the set of values that are considered for \textsf{num-w} (resp. \textsf{num-aa}).
Unless stated otherwise, we consider
\begin{align*}
\mathbb{N}_W &= [d; d+32] \\
\mathbb{N}_{AA} &= \{0,1,2\}.
\end{align*}

In the \textit{suff-perf} scenario we design the algorithm parameters such that an ideal quantum computer performs sufficiently well to pass the tests $T_G(n,d)$ while minimizing the number of amplitude amplification iterations needed.
It is expected for small size problems that the scenario corresponds to a baseline implementation of a discrete-time quantum walk algorithm.

The \textit{best-perf} scenario corresponds to the set of parameters $(\textsf{num-w}_\text{best}, \textsf{num-aa}_\text{best}) \in \mathbb{N}_W \times \mathbb{N}_{AA}$ that maximizes the probability of success that can be obtained by an ideal quantum computer. \\

\bmhead{Experimental parameters selection}

For the cycle graph problem, we consider $N_\text{batch}=1$ batch of $N_\text{shots}=5000$ shots per problem instance (i.e. fixed graph size $n$, fixed distance $d$, fixed target node) for theoretical numerical simulations, and $N_\text{batch}=5$ batches of $N_\text{shots}=1000$ shots per problem instance for experiments on real hardware.
All possible target nodes are selected, i.e. two problem instances for each problem size. \\

For the 2D-torus graph problem, we solely conduct numerical simulations as our goal is to illustrate expected performances and resources scaling rather than evaluate real hardware since the success criterion seems too challenging to meet for current hardware.
In particular, we consider $N_\text{batch}=1$ batch of $N_\text{shots}=1000$ shots per problem instance, therefore relaxing the requirements defined for benchmark evaluation on real hardware.
Nevertheless, the results are sufficiently precise to be used as a meaningful proof of concept demonstration.

For each problem size, we consider 10 problem instances where the target node is randomly selected.
This results in performing multiple batches for a same problem instance for small problem sizes.
However, as problem size increases, the number of batches evaluated per target node tends to one as it becomes less likely to sample the same target node twice.
The final computation of the metric, for 2D-torus graphs, is based on the results of the four target nodes with largest mean success probability.

\bmhead{Quantum routines implementation}

As required by the protocol, we implement a discrete-time quantum walk algorithm based on a Hadamard coin and symmetric coin initial state: $SH\ket{0}_c$ for cycle graphs, $SH\otimes I \ket{00}_c$ for 2D-torus graphs.
We choose to construct the shift operator circuit from CNOT-based increment/decrement functions~\cite{douglas2009efficient,georgopoulos2021comparison,nandi2025robust} for simplicity and resource management.

When combined to the amplitude amplification routine, the output state of DTQW is reflected about the target node's state.

No error mitigation techniques are introduced in this demonstration.

\subsection{Results and discussion}

We first conduct numerical experiments corresponding to the \textit{suff-perf} and \textit{best-perf} scenarios on a noiseless quantum computer simulated by qiskit's AerSimulator backend in order to estimate ideal performances, identify corresponding parameters value, and study the scaling of required algorithm resources.

The following results are obtained for both cycle and 2D-torus graph problems, and should be considered as proof-of-concept for the QWS protocol demonstration.

\bmhead{Numerical simulations and resource estimation}

\begin{figure}[H]
	\begin{subfigure}{.49\textwidth}
		\centering
		\includegraphics[width=0.99\linewidth]{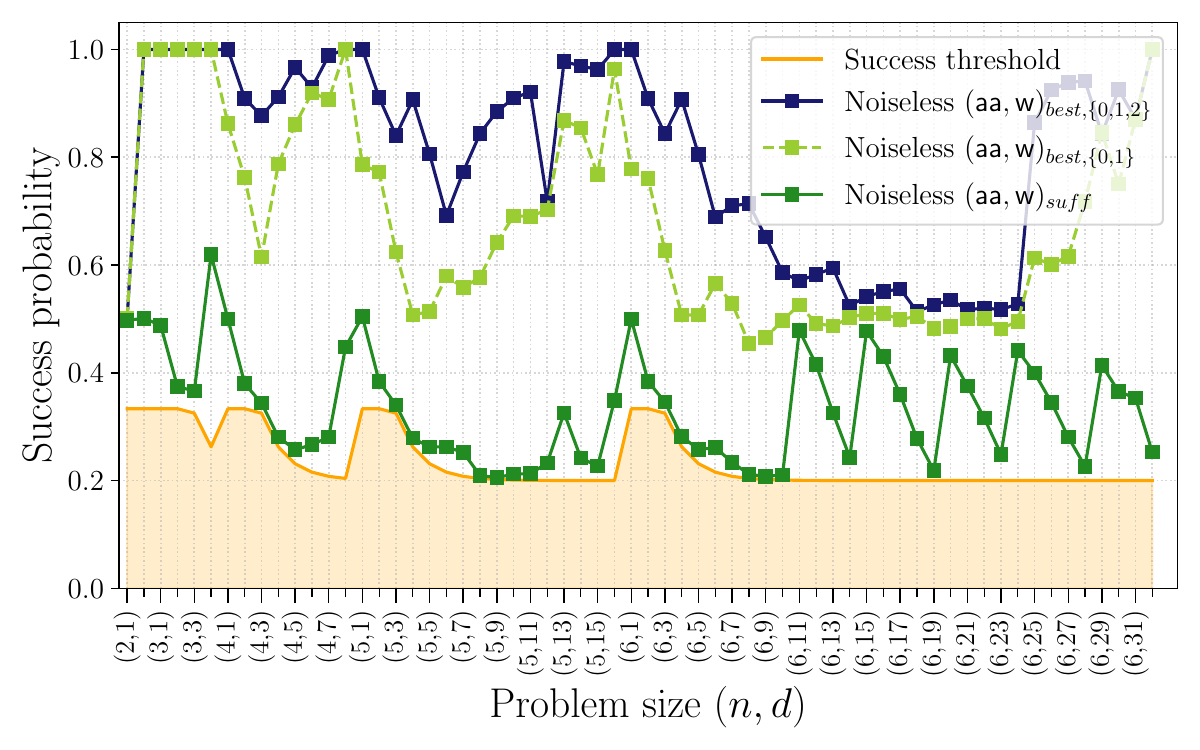}
		\caption{QWS$_\text{cycle}$ performance results.}
		\label{fig:perfAA012}
	\end{subfigure}
	\\
	\begin{subfigure}{.49\textwidth}
		\centering
		\includegraphics[width=0.99\linewidth]{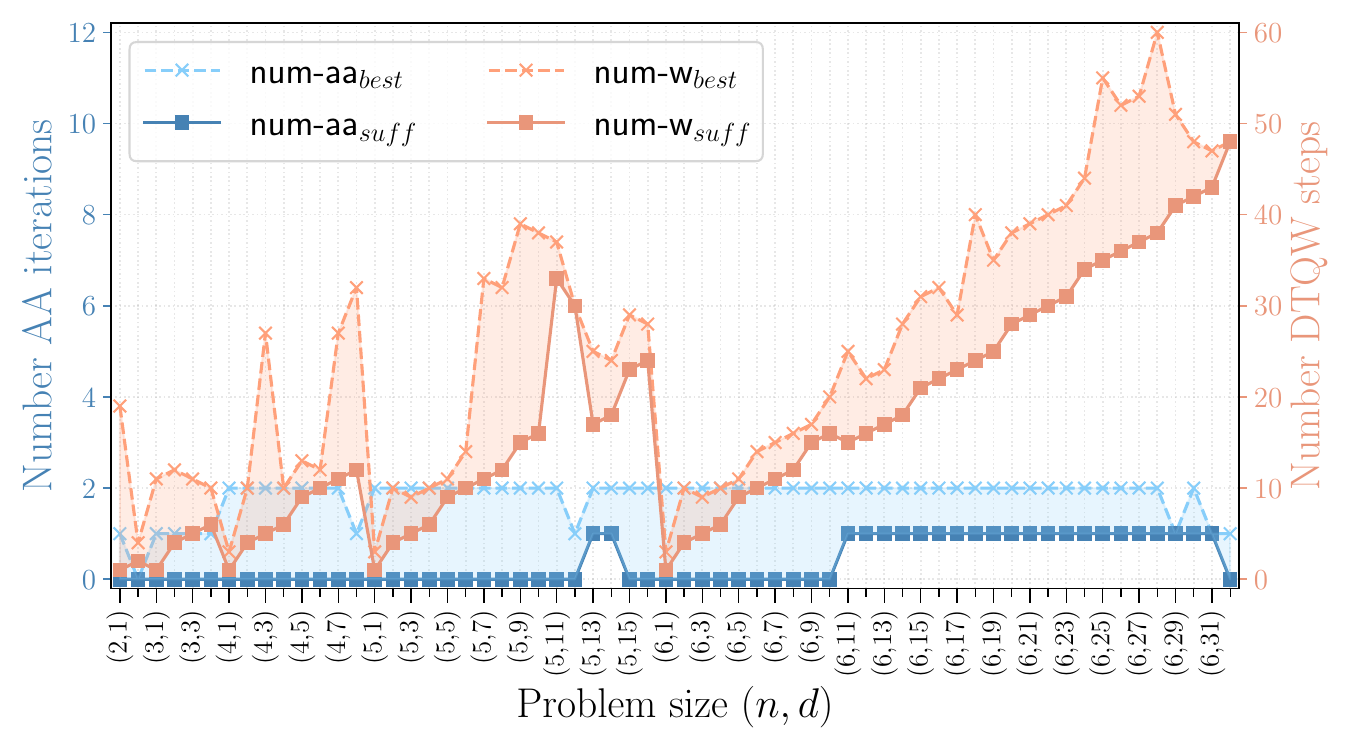}
		\caption{Scaling of algorithm parameters with $\mathbb{N}_{AA}=\{0,1,2\}$, $\mathbb{N}_{W}=[d;d+32]$.}
		\label{fig:resourcesAA012}
	\end{subfigure}
	\caption{QWS$_\text{cycle}$ performance results (a) and algorithm resources (b) under $\textit{suff-perf}$ and $\textit{best-perf}$ scenarios on noiseless QPU.}
\end{figure}

Fig~\ref{fig:perfAA012} represents the success probabilities achieved for the cycle graph in three configurations up to problem size $(n,d)=(6,32)$.
Although all problem sizes are represented on the x-axis, only half of them are labelled for clarity purposes.
The dark green line corresponds to the aforementioned \textit{suff-perf} scenario, while the light green and blue lines correspond to the \textit{best-perf} scenario by considering $\mathbb{N}_{AA}=\{0,1\}$ and $\mathbb{N}_{AA}=\{0,1,2\}$ respectively.

Results from the \textit{suff-perf} scenario highlight the algorithmic possibility to obtain a successful test for all studied problem sizes (i.e. $\text{QWS}_\text{cycle}\geq 64$).
As expected from the definition of the \textit{suff-perf} scenario, the resulting success probabilities remain close to the threshold probability $p^*(n,d)$ while a larger gap is observed for small-distance problem sizes.
We can observe that a high-enough success probability can be obtained by solely relying on the quantum walk routine up to problem size $(5,12)$, for well-selected numbers of steps.
The corresponding values of parameters \textsf{num-w} and \textsf{num-aa} are represented by the full lines in Fig~\ref{fig:resourcesAA012}.

When the problem size increases, however, the numerical experiment confirms that at least one amplitude amplification iteration is necessary to successfully meet the protocol test criterion.
The oscillating behavior of the success probability starting from problem size $(6,11)$ is a consequence of this routine and the choice of scenario.
The value of \textsf{num-aa} is not the optimal value defined in the original paper \cite{brassard2000quantum} that would result in a success probability close to 1 but it is chosen as the minimum requirement to satisfy the benchmark criterion.

In comparison, the \textit{best-perf} scenario amplifies success probabilities up to 1 for certain problem sizes at the cost of larger values of \textsf{num-aa}.
Increasing the size of the set of studied values $\mathbb{N}_{AA}$ would enable to identify \textsf{num-aa} values that significantly enhance the success probability for any problem size.

Furthermore, Fig~\ref{fig:resourcesAA012} highlights that the number of DTQW steps for the \textit{suff-perf} scenario scales linearly with the distance $d$, while the required number of AA iterations scales sub-linearly with respect to $d$.
In the context of \textit{best-perf} scenario, $\textsf{num-aa}_\text{best}$ saturates to $2$, the largest value considered here.
Due to the limited size of $\mathbb{N}_{AA}$, we cannot conclude on the scaling of $\textsf{num-aa}_\text{best}$.
However, $\textsf{num-w}_\text{best}$ is expected to scale linearly with $d$ since the increase in success probability is mostly defined by $\text{num-aa}_\text{best}$.

\begin{figure}[H]
	\begin{subfigure}{.49\textwidth}
		\centering
		\includegraphics[width=0.99\linewidth]{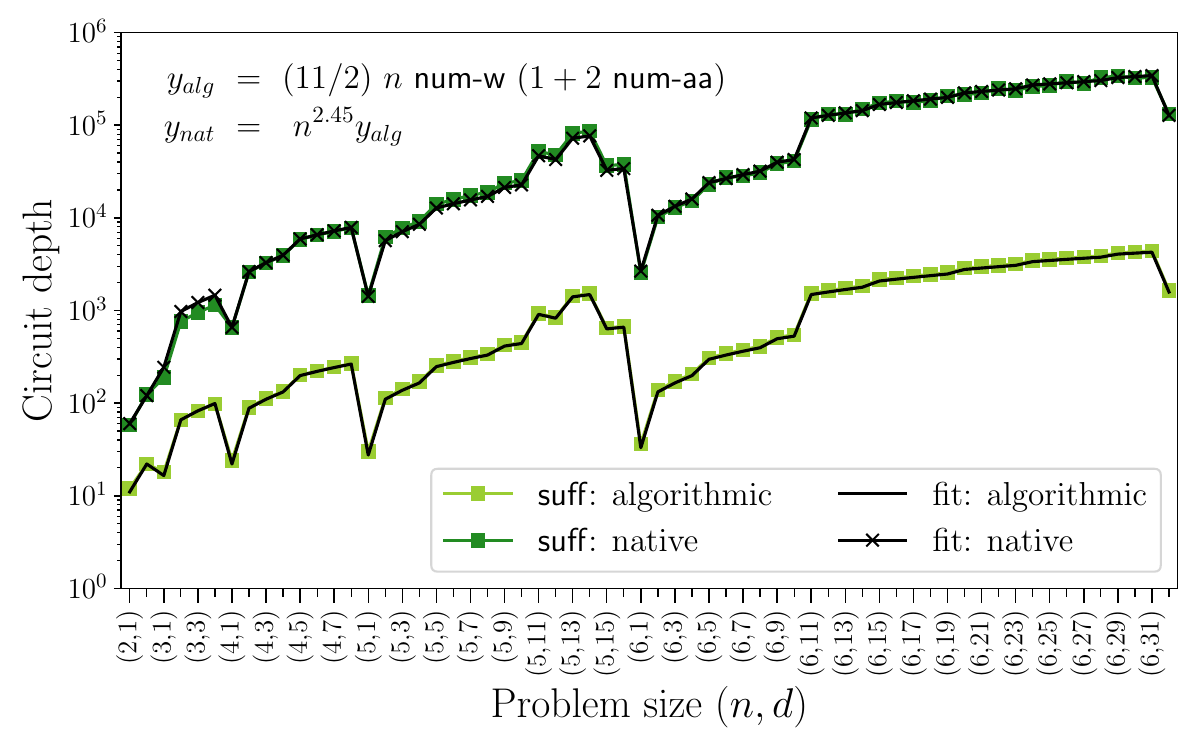}
		\caption{Circuit scaling for the \textit{suff-perf} scenario.}
		\label{fig:depthsuff}
	\end{subfigure}
	\\
	\begin{subfigure}{.49\textwidth}
		\centering
		\includegraphics[width=0.99\linewidth]{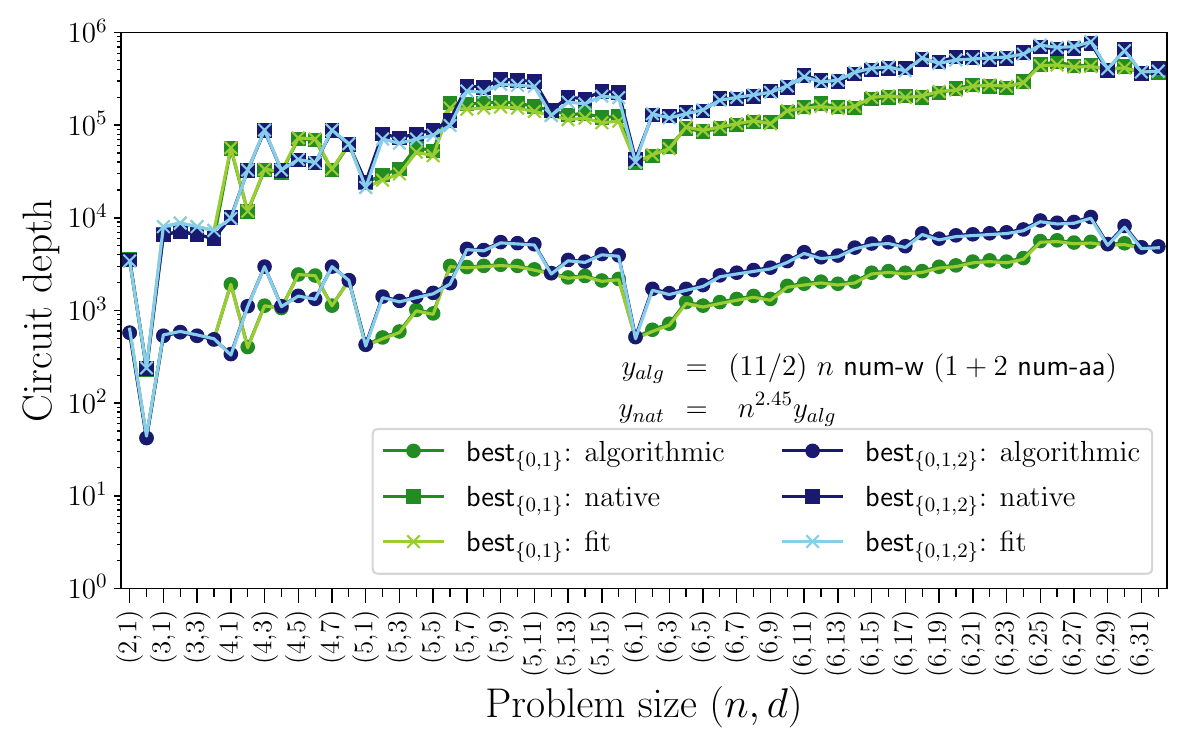}
		\caption{Circuit scaling for the \textit{best-perf} scenario.}
		\label{fig:depthbest}
	\end{subfigure}
	\caption{Algorithmic and native implementation complexity scaling under \textit{suff-perf} (a) and \textit{best-perf} (b) scenarios for cycle graphs on IBM Heron QPU.}
\end{figure}

On a practical point of view, the choice of algorithm parameters strongly affects the depth of the resulting quantum circuit as depicted in Figs~\ref{fig:depthsuff} and~\ref{fig:depthbest} for cycle graphs.
The scaling of the circuit depth can be modeled by the following equations:
\begin{align}
	\text{depth}_\text{alg} &= A \ n \ \textsf{num-w} \ (1+2\textsf{num-aa}) \\
	\text{depth}_\text{nat} &= A \ n^{3.45} \ \textsf{num-w} \ (1+2\textsf{num-aa})
\end{align}
where, $\text{depth}_\text{alg}$ is the theoretical algorithmic circuit depth from the hardware-independent circuit, $\text{depth}_\text{nat}$ the depth of the transpiled circuit based on the native gate set of IBM's Heron architectures, and $A\in\mathbb{R}_{+}^{*}$ a constant.
From the fit equations, we can derive that the theoretical circuit depth for a successful test typically scales as $O(nd)$ while the implemented circuit, which is hardware-dependent, scales as $O(n^{3.45}d)$ for Heron architectures.

The addition of AA iterations results in a multiplicative factor in the number of operations, as it can be seen in the \textit{suff-perf} scenario for problem sizes $(6,11)$ to $(6,31)$.
This corresponds to the design of this routine, which requires to repeat the initial algorithm (i.e. DTQW steps) and its inverse multiple times before measuring the output state.

The circuit depth for the \textit{best-perf} scenario seems to increase slowly with the problem size.
This trend can be explained by the fact that amplitude amplification is performed on almost all problem sizes with a quasi-constant value $\textsf{num-aa}_\text{best} = 2$.
The impact of the distance $d$ on the number of operations is therefore significantly reduced in comparison to the impact of \textsf{num-aa}.
Importantly, we consider a restricted set $\mathbb{N}_{AA}$ in this proof-of-concept.
Increasing the size of this set would likely highlight a variation in the value of \textsf{num-aa} that would be reflected to a larger extent in the resource estimation scaling for this scenario. \\

\begin{figure}[H]
	\begin{subfigure}{.49\textwidth}
		\centering
		\includegraphics[width=0.99\linewidth]{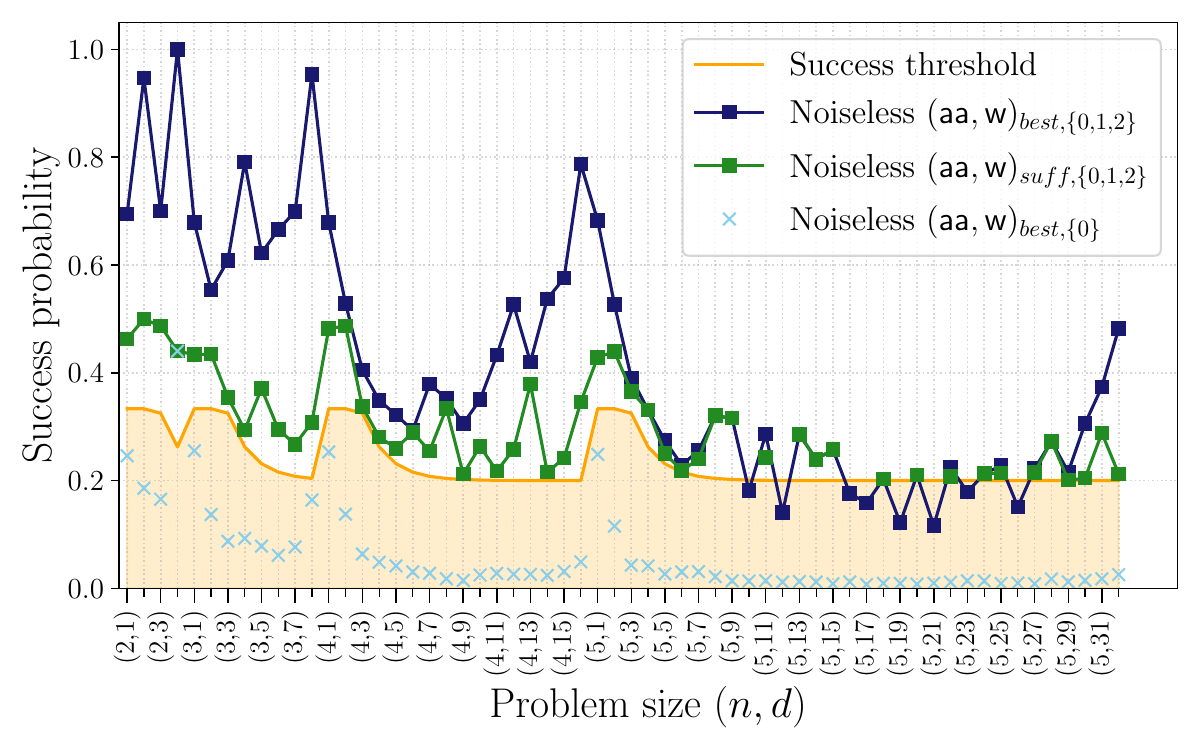}
		\caption{QWS$_\text{torus}$ performance results.}
		\label{fig:torusperf}
	\end{subfigure}
	\\
	\begin{subfigure}{.49\textwidth}
		\centering
		\includegraphics[width=0.99\linewidth]{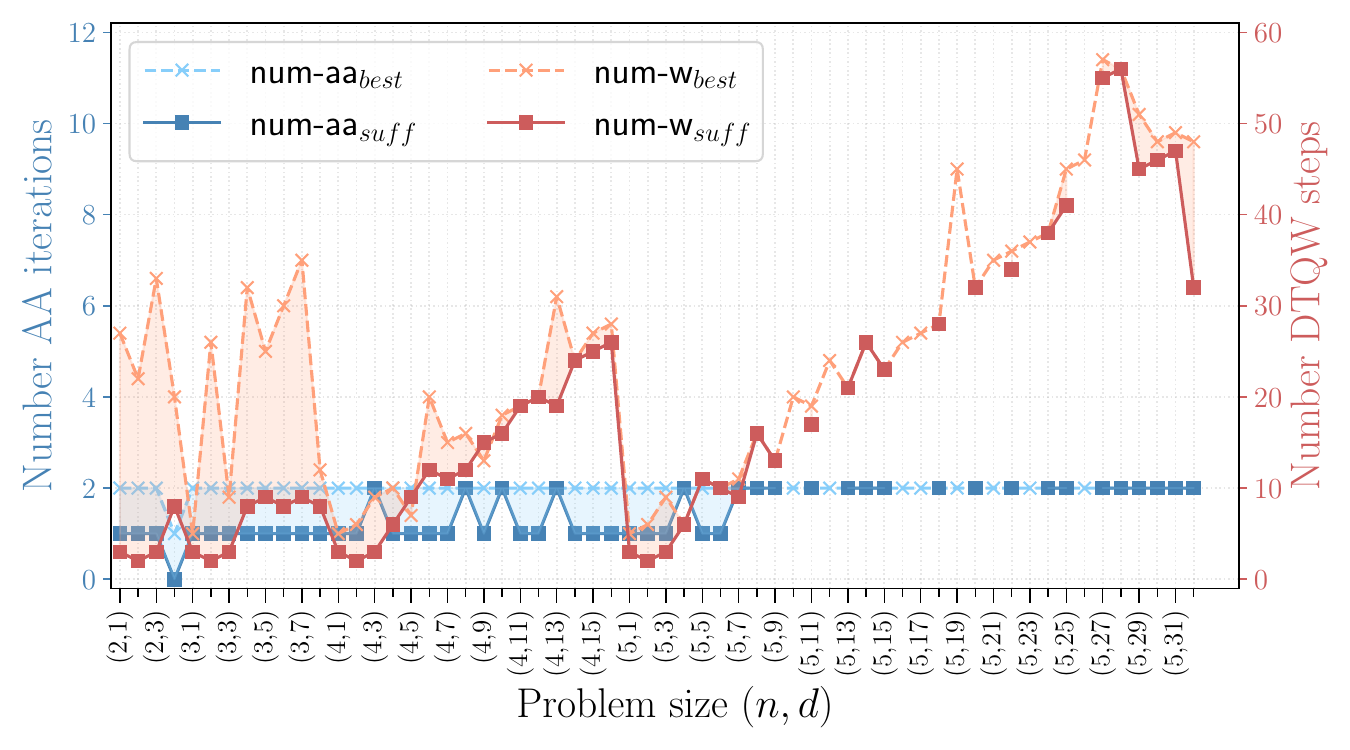}
		\caption{Scaling of algorithm parameters with $\mathbb{N}_{AA}=\{0,1,2\}$, $\mathbb{N}_{W}=[d;d+32]$.}
		\label{fig:torusresources}
	\end{subfigure}
	\caption{QWS$_\text{torus}$ performance results (a) and algorithm resources (b) under $\textit{suff-perf}$ and $\textit{best-perf}$ scenarios on noiseless QPU.}
\end{figure}

The results of numerical simulations for QWS on the 2D-torus graph problem are presented in Fig~\ref{fig:torusperf} and Fig~\ref{fig:torusresources} to highlight expected performance of a noiseless QPU in the proposed scenarios.

On the one hand, we observe that the success criterion can in principle be met with resonable parameter values for a large set of the studied problem sizes, similarly to the cycle graph case.
Both scenarios fail for the first time for a problem size of $(5,10)$, i.e. $\text{QWS}_\text{torus}=41$.
Although the number of AA iterations is upper bounded by 2, the \textit{best-perf} scenario guarantees successful tests with a reasonable margin for error as $p(n,d)\geq 1.2 p^*(n,d)$ up to $(5,3)$.
The scaling of parameter values showcased in Fig~\ref{fig:torusresources} seems analogous to that for the cycle graph: \textsf{num-w} increasing linearly with $d$.
Nevertheless, the set $\mathbb{N}_{AA}$ needs to be further extended to allow for successful tests on large problem sizes, and help estimate the scaling of \textsf{num-aa}.

On the other hand, when no amplitude amplification is considered (light blue crosses), we observe that the success probability remains below the minimum threshold for all problem sizes except size $(2,4)$, i.e. $\text{QWS}_\text{torus}=0$.
This phenomenon occurs despite the large range of values evaluated for \textsf{num-w} and the advantageous post-selection of the four ideal target nodes to consider for metric calculation.
It highlights the robustness of the protocol with respect to target nodes finetuned-selection and its limited impact on the final score. \\

\begin{figure}[H]
	\begin{subfigure}{.49\textwidth}
		\centering
		\includegraphics[width=0.99\linewidth]{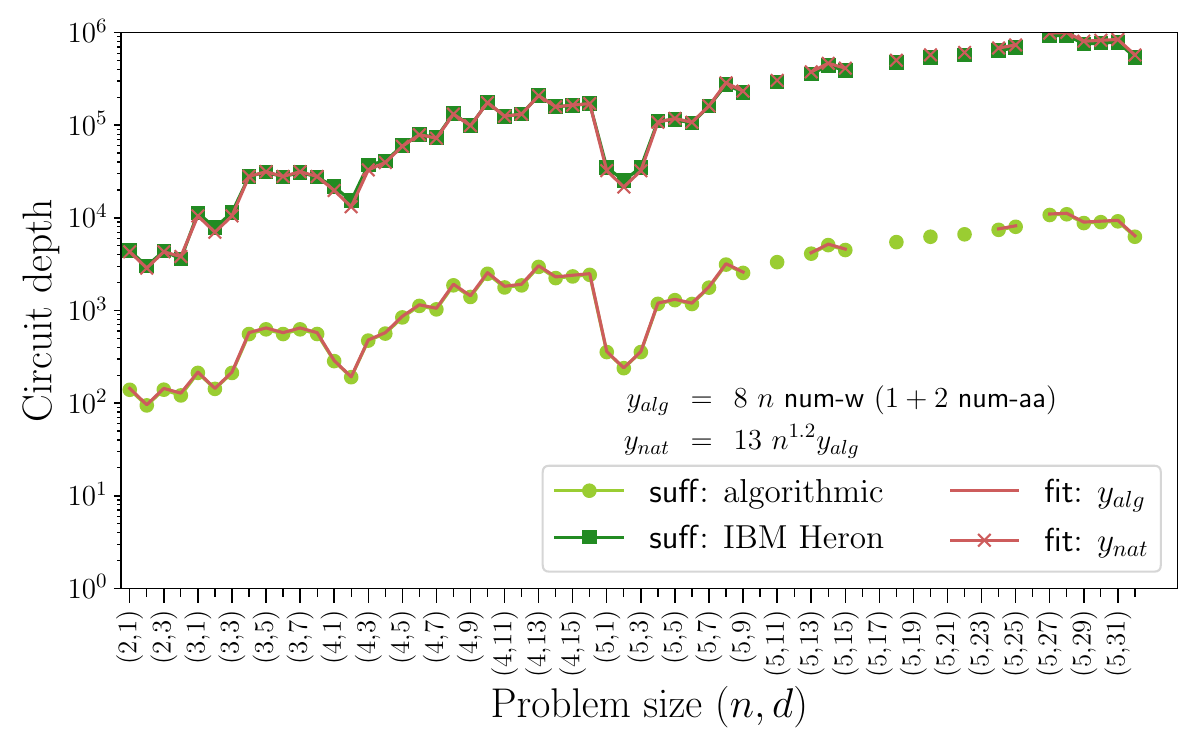}
		\caption{Circuit scaling for the \textit{suff-perf} scenario.}
		\label{fig:torusdepthsuff}
	\end{subfigure}
	\\
	\begin{subfigure}{.49\textwidth}
		\centering
		\includegraphics[width=0.99\linewidth]{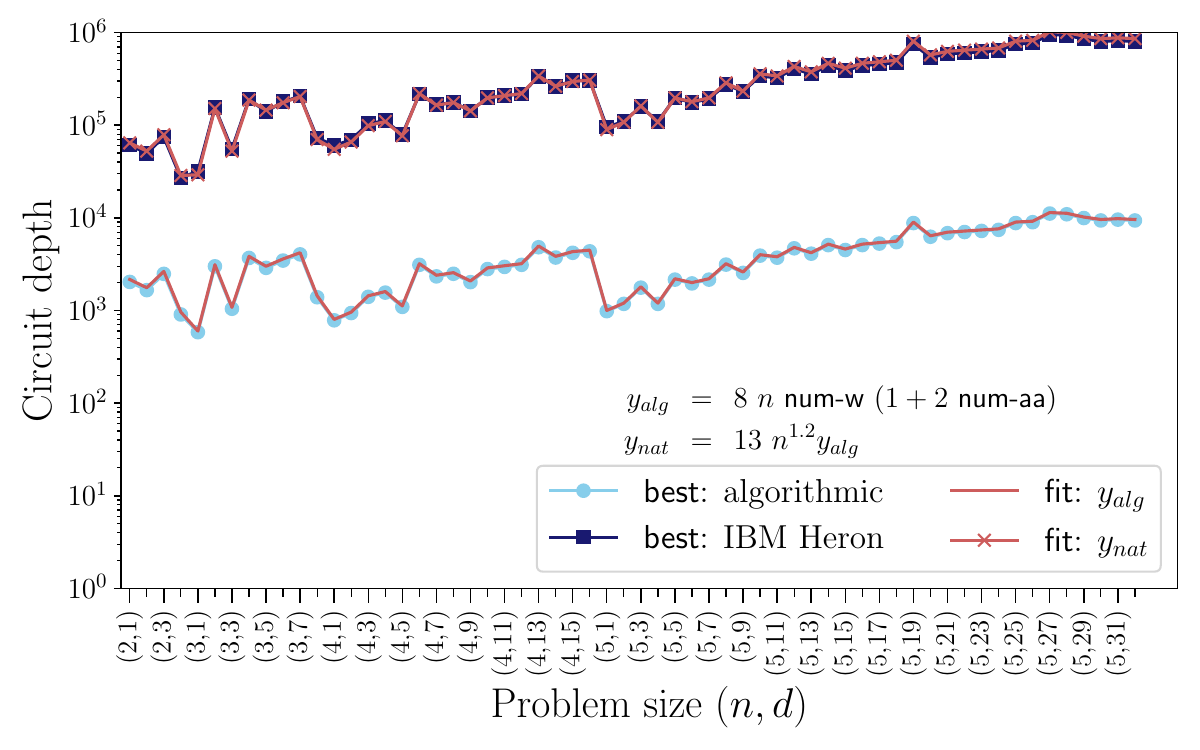}
		\caption{Circuit scaling for the \textit{best-perf} scenario.}
		\label{fig:torusdepthbest}
	\end{subfigure}
	\caption{Algorithmic and native implementation complexity scaling under \textit{suff-perf} (a) and \textit{best-perf} (b) scenarios for 2D-torus graphs on IBM Heron QPU.}
\end{figure}

The evolution of the circuit depth with respect to problem size is represented in Fig~\ref{fig:torusdepthsuff} and Fig~\ref{fig:torusdepthbest}.
The overall scaling of the circuit depth in both scenarios is in adequation with that of the cycle graph problem, up to a multiplicative factor: the theoretical circuit depth evolves as $O(nd)$ (by assuming \textsf{num-aa} varies sublinearly with $d$) while the Heron architecture-dependent circuit depth scales as $O(n^{2.2}d)$ under the same assumptions.
The latter implies that improving the QWS$_\text{torus}$ score can prove to be easier than improving the QWS$_\text{cycle}$ score on a similar range, especially considering that the difference in circuit depth becomes limited starting from a score of 62 in the \textit{suff-perf} scenario.
Although the proposed fit function appears to rightfully match numerical data, the values themselves should be treated carefully for the \textit{best-perf} scenario as some of the largest problems considered are not successfully solved with the studied parameters.
In other words, circuit-depth scaling under the \textit{best-perf} scenario is expected to significantly increase when considering successful parameter values. 

\bmhead{Experiments on IBM's hardware}

\begin{figure}[H]
	\begin{subfigure}{.49\textwidth}
		\centering
		\includegraphics[width=0.99\linewidth]{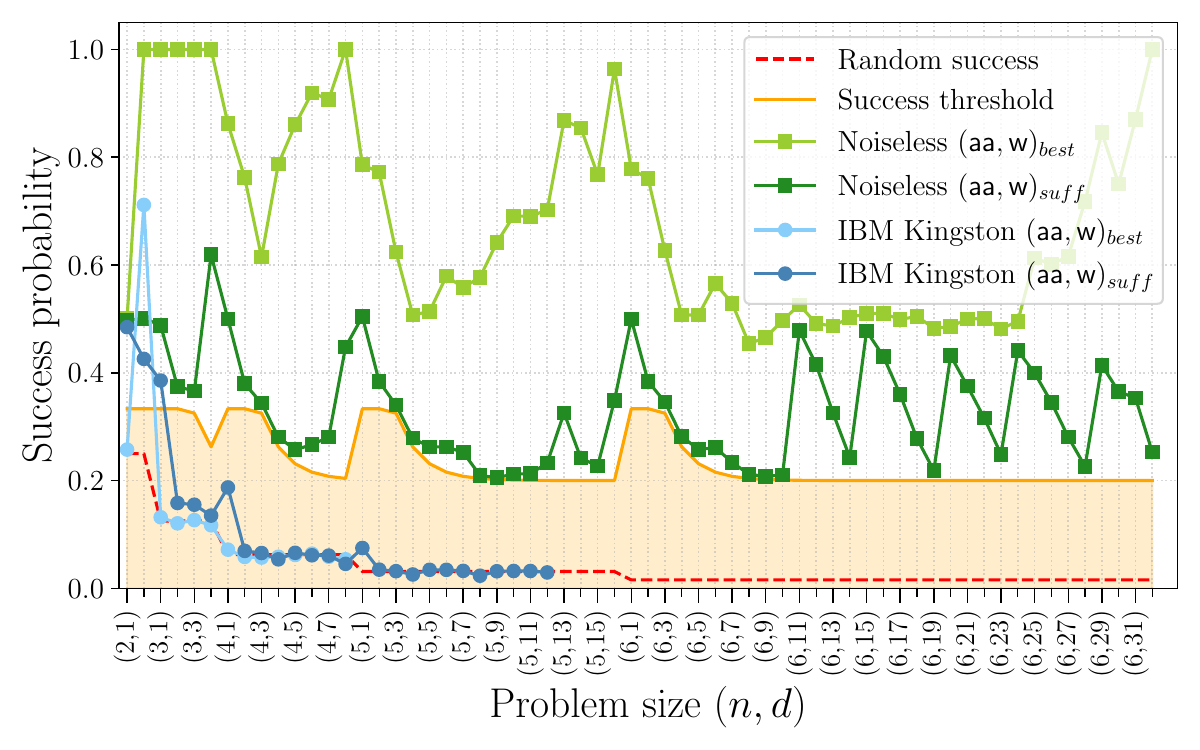}
		\caption{Performance comparison IBM Kingston vs theory.}
		\label{fig:perfkingston}
	\end{subfigure}
	\\
	\begin{subfigure}{.49\textwidth}
		\centering
		\includegraphics[width=0.99\linewidth]{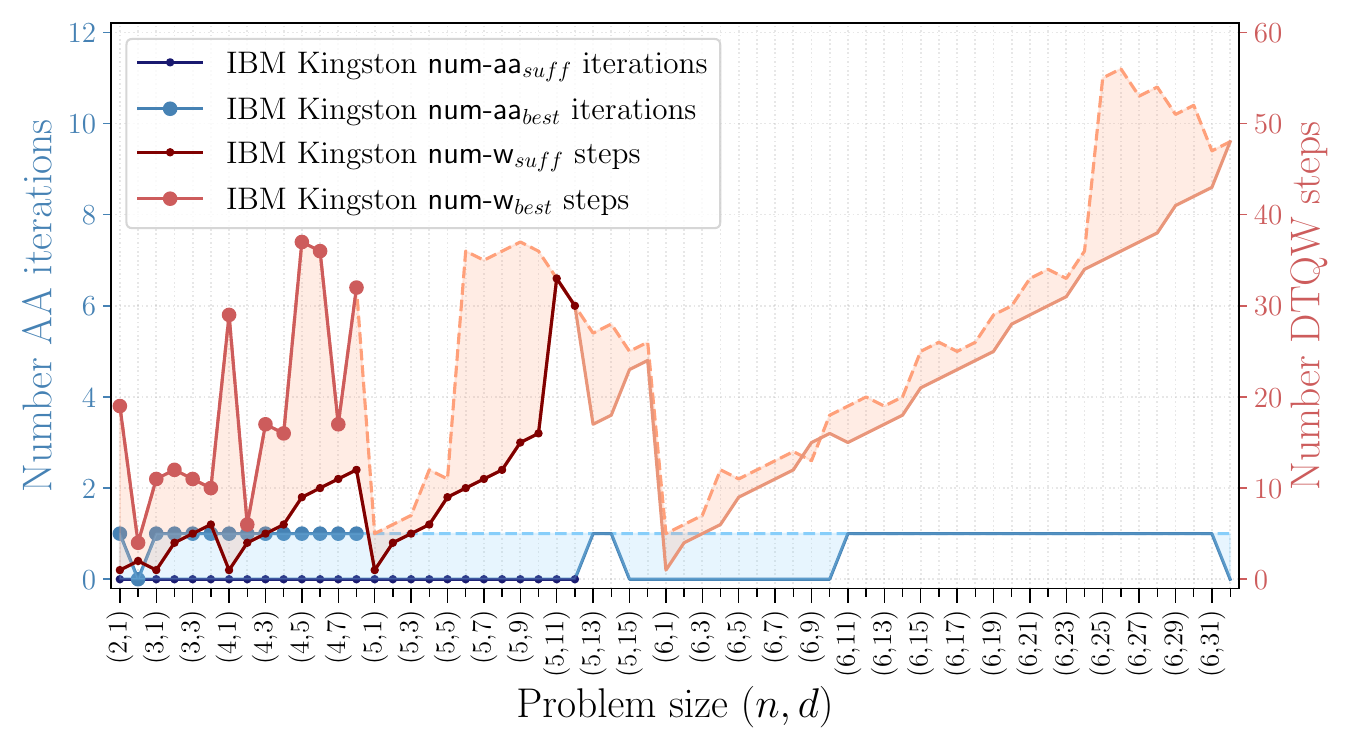}
		\caption{Selected algorithm parameters \textsf{num-w}, \textsf{num-aa}.}
		\label{fig:resourceskingston}
	\end{subfigure}
	\caption{QWS$_\text{torus}$ performance results (a) and algorithm resources (b) under $\textit{suff-perf}$ and $\textit{best-perf}$ scenarios on IBM Kingston (Heron r2).}
\end{figure}

Figs~\ref{fig:perfkingston} and~\ref{fig:resourceskingston} describe the \textit{suff-perf} and \textit{best-perf} scenarios experiments on the cycle graph problem, run on the ibm\_kingston QPU, in comparison to the results on a noiseless simulator.
It can be observed that the success probabilities in the \textit{suff-perf} scenario exceed the threshold probability only up to size $(3,1)$ before plummeting to results similar to random guesses, at the exception of problem size $(4,1)$, resulting in $\text{QWS}_\text{cycle}(\text{ibm\_kingston})=5$.
Although the first three tests are successful, the success probability decreases from $0.5$ to $0.39$ where ideal results show a constant success probability of $0.5$.

This decrease, as well as the overall trend of these experimental results, is an expected behavior due to noise propagation in the deep quantum circuits according to hardware characteristics from Table~\ref{tab:qpu}.

Let us provide an order of magnitude for expected circuit depth manageable on current processors.
The IBM Kingston hardware proposes single and two-qubit median gate times of $t_{1q}=32$ns and $t_{2q}=68$ns.
Assuming an exponential decay in time for the qubits state due to decoherence, and considering IBM Kingston's $T_1$ and $T_2$ characteristics, we expect the QPU to run a circuit of maximum depth $850$ while maintaining $75\%$ of the quantum states intact.
However, errors in gates operations also accumulate and affect circuit runs.
Assuming a repartition $85\slash 15$ for single/two-qubit gates (consistent with our experiments), such a 850-depth circuit would be successfully implemented in practice with a probability of $12\%$.

Correlating these estimates with the resource estimation in Fig~\ref{fig:depthsuff} highlights that running circuits of more than a thousand gates on this hardware results in highly noisy outcomes.
Circuits of about $700$ gates, e.g. problem size $(4,1)$, result in deteriorated results for the Heron r2 architecture (see Fig~\ref{fig:perfkingston}) compared to ideal simulations but the outcome remains consistent with the expected behavior of the quantum computation.

Similarly, in the \textit{best-perf} scenario with $\mathbb{N}_{AA}=\{0,1\}$, results match the behavior of random guesses except for problem size $(2,2)$ where the success probability reaches $0.72$.
The corresponding $300$-depth circuit is particularly narrow as it does not make use of amplitude amplification, hence the high success probability.
Other instances implementations contain more than $3000$ layers of operations, due to additional AA routine iterations, therefore providing completely noisy results on current hardware.
\\

\begin{figure}[H]
	\centering
	\includegraphics[width=0.99\linewidth]{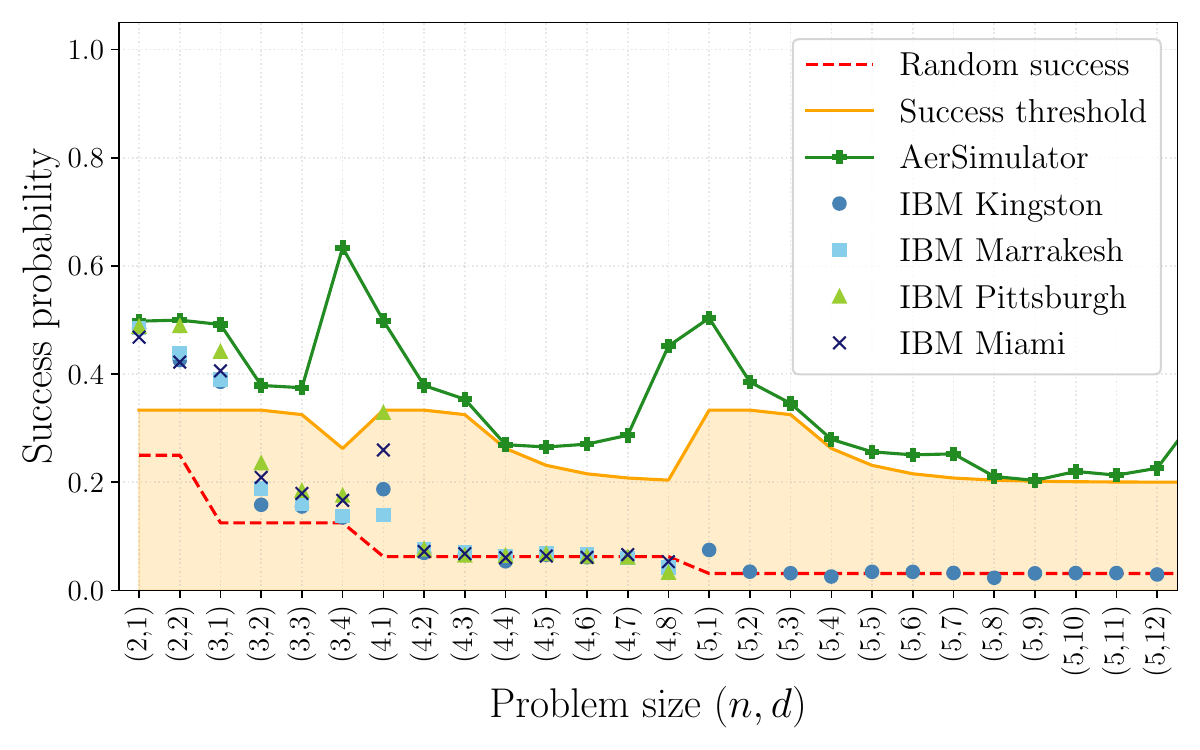}
	\caption{QWS evaluation (cycle graph) on IBM hardware under \textit{suff-perf} scenario.}
	\label{fig:expsuff}
\end{figure}

A comparison of the \textit{suff-perf} results from the different generations of IBM hardware on the QWS benchmark on cycle graphs is proposed in Fig~\ref{fig:expsuff}.
Experiments are run for problem sizes up to $(n,d)=(4,8)$ for the Heron r2, Heron r3, and Nighthawk r1 QPU.
We observe the same trend as for ibm\_kingston Heron r2 experiments due to relatively similar error rates and coherence times.
Nevertheless, the ibm\_pittsburgh Heron r3 hardware provides slightly more accurate success probabilities even though $1000$-depth circuits remain too challenging for all current architectures.
The value of the Quantum WalkScore would be $\text{QWS}_\text{cycle}=5$ for all these hardware with the selected \textit{suff-perf} parameters.

As QPU technology matures, the number of problems that can be successfully solved (and thus the QWS) is expected to increase according to the scaling described in Fig~\ref{fig:depthsuff} and Fig~\ref{fig:torusdepthsuff}.

\section{Conclusion}\label{sec:conclusion}

In this work, we have introduced Quantum WalkScore (QWS), a scalable application-oriented benchmark protocol to evaluate the performance of current and future gate-based quantum computers.
QWS measures how effectively hardware executes two core quantum primitives --discrete-time quantum walks and quantum amplitude amplification-- as building blocks for graph nodefinding tasks.
By focusing on graph-structured abstraction, QWS connects hardware performance to the achieved success probability for finding a designated target node on a given graph topology.
This abstraction is directly relevant to problems that can be modeled as marked-node finding on graphs, including database search, network navigation, or threat localization tasks for instance.

For practical and fair benchmarking, we have studied cycle and 2D-torus graphs. These structures enable controlled scalability and allow QWS dicriminate performance across current noisy quantum processors and fault-tolerant quantum computers.
We have illustrated the QWS protocol with two example parameter-selection scenarios, showing that the achieved QWS score can be significantly affected by the choice of algorithm design parameters \textsf{num-w} and \textsf{num-aa}.
Finally, we have demonstrated practical evaluation of QWS through experiments on IBM's commercially available Heron r2, Heron r3, and Nighthawk r1 quantum processors, achieving a proof-of-concept $\text{QWS}_\text{cycle}$ score of $5$ for the tested hardware under the \textit{suff-perf} scenario.
The results highlight the technological challenge of executing the combined DTQW-AA circuits required by QWS, whose circuit depth scales polynomially with $n$ and $d$ in our construction.
Overall, we believe that this interplay of algorithmic and hardware challenges makes QWS a suitable basis for application-driven performance evaluation of quantum computers.

\bmhead{Acknowledgements}

As part of the MetriQs-France program, this work is supported by France 2030 under the French National Research Agency grant number ANR-22-QMET-0002. We acknowledge the use of IBM Qiskit services for this work. The views expressed are those of the authors and do not reflect the official policy or position of IBM  or the IBM Quantum team.

\bibliography{biblio}

\end{document}